\documentclass[
  prx,
  twocolumn,
  reprint,
  superscriptaddress,
  amsmath,amssymb,
  aps,
  floatfix,
  longbibliography,
  nofootinbib,
]{revtex4-2}

\usepackage{graphicx}
\graphicspath{{figures/}}
\usepackage[caption=false]{subfig}
\usepackage{bm}
\usepackage{braket}
\usepackage[dvipsnames]{xcolor}
\definecolor{ibmblue}{HTML}{0F62FE}
\usepackage[
    colorlinks=true,
    citecolor=ibmblue,
    urlcolor=ibmblue,
    linkcolor=ibmblue
]{hyperref}
\usepackage[capitalise]{cleveref}

\usepackage{booktabs}
\usepackage[version=4]{mhchem}

\usepackage[T1]{fontenc}

\usepackage[ruled,vlined,linesnumbered,noend]{algorithm2e}
\crefname{algocf}{Algorithm}{Algorithms}
\Crefname{algocf}{Algorithm}{Algorithms}
\SetKwComment{Comment}{\hfill$\triangleright$~}{}
\SetCommentSty{textit}

\usepackage{siunitx}
\DeclareSIUnit\hartree{Ha}
\DeclareSIUnit\kcal{kcal}
\DeclareSIUnit\angstrom{\text {Å}}

\newcommand{\oxoMn}{oxo-\ce{Mn}(Salen)}

\newcommand{\IBMZRL}{IBM Research}
\newcommand{\EPFL}{Institute of Physics, École Polytechnique Fédérale de Lausanne (EPFL), CH-1015 Lausanne, Switzerland}
\newcommand{\ETH}{Institute for Theoretical Physics, ETH Zürich, CH-8093 Zürich, Switzerland}

\begin{document}

\title{Spin-adapted sample-based quantum diagonalization with randomized time evolution}

\author{Stefano Barison}
\email{stefano.barison1@ibm.com}
\affiliation{\IBMZRL}%
\affiliation{\ETH}
\author{Samuele Piccinelli}
\affiliation{\IBMZRL}%
\affiliation{\EPFL}%
\author{Alberto Baiardi}
\affiliation{\IBMZRL}%
\author{Francesco Tacchino}
\affiliation{\IBMZRL}%
\author{Ivano Tavernelli}
\affiliation{\IBMZRL}%

\begin{abstract}
Quantum algorithms based on the classical post-processing of measurement samples have emerged as effective and robust methods to approximate low-energy wavefunctions of many-body quantum systems on present-day and early-fault-tolerant quantum devices.
This paper introduces a generic framework for enforcing spin symmetry within these approaches.
We tailor this framework to SqDRIFT, a sample-based quantum diagonalization variant that uses the qDRIFT randomized compilation protocol to generate the quantum samples.
We discuss how three key components of SqDRIFT---the initial-state preparation, the construction of the time-evolution circuits, and the classical post-processing---can be adapted to incorporate spin symmetry.
We show that spin-pure circuits alone are not sufficient, since the sampled determinants generally span a subspace that is not closed under $\hat{S}^2$, and we restore this closure by completing the sampled electronic configurations.
In classical simulations of 4-aminobenzonitrile and of an \ce{FeN4} model complex, and in experiments on an IBM quantum processor for the \oxoMn{} complex, this spin-adapted post-processing yields roots that are exact spin eigenfunctions at every subspace size.
On hardware, it does so with diagonalized subspaces two to three orders of magnitude smaller than those of the conventional constructions for the same circuit depth.
\end{abstract}

\maketitle

\section{Introduction}
\label{sec:introduction}

Solving the ground-state problem for quantum-chemical Hamiltonians is one of the most promising applications of quantum computers.
Despite that, realizing this promise in practice, on present-day quantum computers, has proven challenging.
While approaches based on the variational quantum eigensolver can hardly be scaled to systems with more than 20 electrons in 20 orbitals~\cite{Wecker2015}, methods based on quantum simulation primitives, such as quantum phase estimation and quantum subspace approaches, yield very deep circuits due to the complexity of implementing the time-evolution operator~\cite{Lee2021, Goings2022}.
Sample-based approaches, such as quantum selected configuration interaction~\cite{Kanno2026,Sugisaki_2025, Mikkelsen_2025} and sample-based quantum diagonalization (SQD)~\cite{RobledoMoreno2025,Barison2025,shirakawa2025,Kaliakin2025,merz2026}, have recently emerged as an intermediate solution for executing large-scale chemical simulations on current quantum hardware.
The key idea underlying these approaches is to solve the ground-state problem classically within a subspace that is constructed by sampling from a (collection of) quantum circuit(s).
Sample-based methods in which circuits are constructed from time-evolution primitives, such as sample-based Krylov quantum diagonalization (SKQD)~\cite{Yu2025}, are especially appealing: under the assumption that the ground state is concentrated on a small (polynomially scaling) number of determinants, they inherit convergence guarantees similar to those of quantum phase estimation.
Since the exact time-evolution operator is too costly to implement on present-day quantum computers, it must be approximated when applying these approaches to chemical Hamiltonians.
In SqDRIFT~\cite{Piccinelli2025}, the time-evolution operator is approximated through the qDRIFT randomized protocol~\cite{Campbell2019}.
Early applications of SqDRIFT to large-scale quantum-chemical calculations~\cite{Roncevic2026,Piccinelli2026} have demonstrated the potential of the method.

In this work, we further extend SqDRIFT to include spin symmetry.
From a practical perspective, accounting for spin symmetry is a critical requirement of any method aimed at simulating strongly correlated systems.
In fact, many strongly correlated systems feature non-trivial spin energetics~\cite{Sharma2014,Li2019,LiManni2018_FeP}, which can be resolved only using electronic-structure methods properly accounting for this symmetry.
While spin-adapted variants of many state-of-the-art classical approaches for simulating strongly correlated systems are nowadays available, including tensor networks~\cite{Sharma2012,Keller2016,Schmoll2020} and selected configuration interaction (SCI)~\cite{Dobrautz2019,LiManni2020,Chilkuri_2021,Ugandi2023}, this is not yet the case for most sample-based quantum approaches, and in particular for those based on time evolution.
This work fills this gap and discusses how the three key components underlying SqDRIFT, i.e., the initial-state preparation~\cite{Carbone2022_SpinEigenfunctions,Tew2025_SpinCoupling,Ollitrault2024_OrbitalOpt}, the qDRIFT randomization protocol~\cite{Campbell2019}, and the classical post-processing can be modified to account for the Hamiltonian spin symmetry.
Specifically, we discuss how to express the second-quantized form of the electronic-structure Hamiltonian in a spin-invariant way, such that each individual term that is sampled in qDRIFT commutes with $\hat{S}^2$.
Moreover, we discuss how to modify the classical post-processing originally proposed in Ref.~\cite{RobledoMoreno2025} to ensure that the diagonalization subspace is closed under $\hat{S}^2$, so that the wavefunctions obtained through subspace diagonalization are eigenfunctions of $\hat{S}^2$.
Our proposed procedure follows the approaches presented in Refs.~\cite{Chilkuri_2021,Sugisaki_2025}, which we combine with configuration recovery~\cite{RobledoMoreno2025}. 
We validate the approach through classical simulations of the excited states of 4-aminobenzonitrile (ABN) and of the quintet--triplet splitting of the \ce{[Fe(NH3)4]^{2+}} model complex (which we will refer to as \ce{FeN4}).
Our extension enables applying the pipeline to data obtained from quantum hardware, as we show with experiments on an IBM Heron processor targeting the spin energetics of the \oxoMn{} complex, a transition-metal complex that has been extensively studied with tensor-network approaches~\cite{Wouters2014,OlivaresAmaya2015,Stein2016_AutoCAS}.
The \ce{FeN4} simulations show that spin-adapted circuits alone do not yield spin-pure eigenstates of the subspace Hamiltonian, since the sampled determinants span a subspace that is not closed under $\hat{S}^2$, whereas spin completion does at every subspace size.
On hardware, where only spin-adapted post-processing is applied, spin completion yields exact spin eigenfunctions at comparable accuracy, while requiring diagonalization of subspaces two to three orders of magnitude smaller than those required by previous construction strategies at the same excitation level.
Our work lays the groundwork for constructing an SqDRIFT-based pipeline that can target strongly correlated systems with non-trivial spin energetics~\cite{Sharma2014,Bogdanov2022_Cuprates}.

\section{Theory}
\label{sec:theory}

We focus on the non-relativistic electronic-structure Hamiltonian formulated in second quantization within a finite orbital basis, given by:
\begin{equation}
 \begin{aligned}
  \hat{H}_\text{e} &= 
    \sum_{p,q=0}^{N_o-1} \sum_{\sigma \in \left\{ \alpha, \beta \right\}}
        h_{pq} \hat{a}_{p \sigma}^\dagger \hat{a}_{q \sigma} \\
    &+ \frac{1}{2} \sum_{p,q,r,s=0}^{N_o-1}
      \sum_{\sigma, \sigma^\prime \in \left\{ \alpha, \beta \right\}}
        g_{pqrs} \hat{a}_{p \sigma}^\dagger \hat{a}_{r \sigma^\prime}^\dagger \hat{a}_{s \sigma^\prime} \hat{a}_{q \sigma} \, ,
 \end{aligned}
 \label{eq:mol_hami}
\end{equation}
where $h_{pq}$ and $g_{pqrs}$ are one- and two-electron integrals, respectively, and $\hat{a}_{p \sigma}^\dagger$ ($\hat{a}_{p \sigma}$) is the creation (annihilation) operator of the fermionic mode $p$ with spin $\sigma$.
In the following, we first describe how to modify the SqDRIFT pipeline to ensure that the circuits prepare spin-pure wave functions.
Then, we describe how to generalize the classical post-processing of quantum samples to enforce spin completeness of the sampled subspaces.

We first define \textit{determinants} and \textit{electronic configurations}, since these terms are often used interchangeably in the literature, whereas here they have distinct meanings.
Given an ordered orbital basis set, we refer to electronic configuration (or, simply, configuration) as a vector of $N_o$ integers whose $i$-th entry is 0, 1, or 2 if orbital $i$ is empty, singly occupied, or doubly occupied, respectively.
Note that, in an electronic configuration, the information about singly occupied orbitals is not spin-resolved.
Conversely, \textit{determinants} here refers to Slater determinants
\begin{equation}
\label{eq:slat_det}
    \ket{\bm{x}} = \prod_{\sigma = \alpha, \beta} \, \prod_{i=0}^{N_o-1} \left[ \hat{a}^{\dagger}_{i \sigma}\right]^{x_{i \sigma}} \ket{\varnothing} \, ,
\end{equation}
where $\ket{\varnothing}$ is the vacuum state and $x_{i \sigma} \in \{0,1\}$. The determinant is thus identified by a bitstring $\bm{x} = (\bm{a}, \bm{b})$ of length $2N_o$, where $\bm{a}, \bm{b} \in \{0,1\}^{N_o}$ are the occupation strings of the $\alpha$ and $\beta$ spin orbitals, respectively, and we write $\ket{\bm{x}} = \ket{\bm{a},\bm{b}}$.
Slater determinants form an orthonormal basis of the antisymmetric (fermionic) sector of the many-particle Hilbert space.
Hence, any molecular wavefunction can be expressed as
\begin{equation}
    \label{eq:slater_wf}
    \ket{\psi} = \sum_i c_i \ket{\bm{x}_i} \,, \quad \text{with } c_i \in \mathbb{R}  \, .
\end{equation}
In a Slater determinant, every electron has a well-defined spin.
Hence, Slater determinants are eigenfunctions of $\hat{S}_z$, with eigenvalue $M_S = (N_\alpha - N_\beta)/2$, where $N_\alpha$ ($N_\beta$) is the number of $\alpha$ ($\beta$) electrons, but in general not of the total spin $\hat{S}^2$.
However, the non-relativistic electronic structure Hamiltonian in \cref{eq:mol_hami} fulfills the following commutation relations:
\begin{equation}
    \label{eq:hami_commutation}
    \left[ \hat{H}_\text{e} \, , \, \hat{S}_z \right] = \left[ \hat{H}_\text{e} \, , \, \hat{S}^2 \right] = 0 \,.
\end{equation}
Hence, a common eigenbasis of $\hat{H}_\text{e}$, $\hat{S}_z$ and $\hat{S}^2$ exists.
In particular, the eigenstates of $\hat{H}_\text{e}$ can be chosen to be eigenfunctions of the total spin.
While preparing a linear combination of Slater determinants with the same $M_S$ guarantees that the final wavefunction is an $\hat{S}_z$ eigenfunction, this is not true for the total spin since Slater determinants are, in general, not eigenfunctions of $\hat{S}^2$.
Fortunately, we can define linear combinations of Slater determinants that possess the same spatial, spin, and permutational symmetry as the physical eigenstates.
These basis states are called \emph{configuration state functions} (CSFs).
Any molecular wavefunction can be expressed as a linear combination of CSFs
\begin{equation}
    \label{eq:csf_eq}
    \ket{\psi} = \sum_j d_j \, \ket{\text{CSF}_j} \,, \quad \text{with } d_j \in \mathbb{R}  \,.
\end{equation}
Several strategies to construct CSFs exist, such as the graphical unitary group approach (GUGA)~\cite{Shavitt2009_GUGA,Shavitt2009_GUGA-Matrix}, which is widely adopted in (classical) quantum-chemistry software~\cite{Lischka2011_Columbus,LiManni2023_OpenMolcas}.
The GUGA approach has been combined with different variants of modern CI approaches, including selected CI~\cite{Ugandi2023} and full configuration interaction quantum Monte Carlo (FCIQMC)~\cite{Dobrautz2019,Dobrautz2021_GUGA-FCIQMC}.
In the following, we illustrate how CSFs can be used as the initial state for spin-adapted SqDRIFT calculations and how the relationship between CSFs and Slater determinants can be used to efficiently post-process bitstrings sampled from quantum devices.

\subsection{SKQD and SqDRIFT}
\label{sec:theory-sqdrift}

Sample-based approaches diagonalize the Hamiltonian given in \cref{eq:mol_hami} in a subspace defined by the samples collected from a (collection of) quantum circuit(s).
Among the different sample-based quantum diagonalization (SQD) variants, we focus in this work on SqDRIFT, which can be interpreted as a randomized variant of SKQD~\cite{Yu2025}.
SKQD constructs the subspace by sampling from quantum Krylov states, defined as:
\begin{equation}
  \ket{\psi_k} 
    = \mathrm{e}^{-\mathrm{i}\hat{H}_\text{e}k\Delta t} \ket{\psi_0} \, ,
  \label{eq:kqd_circs}
\end{equation}
where the time-evolution operator is compiled using Trotter product formulas.
Conversely, in SqDRIFT, the operator is approximated through the qDRIFT randomized compilation protocol~\cite{Campbell2019}.
To this end, we decompose the Hamiltonian as
\begin{equation}
  \hat{H}_\text{e} = \sum_i w_i h_i\,,
  \label{eq:HamiltonianDecomposition}
\end{equation}
where we intentionally do not specify, for the moment, the form of the operators $h_i$, as it differs between the spin-adapted SqDRIFT variants discussed below.
However, we require, without loss of generality, that $w_i > 0$ and that each $h_i$ has unit spectral norm, $\lVert h_i \rVert = 1$~\cite{Campbell2019}.
The exact time-evolution channel $\mathcal{U}_t[\rho] = \mathrm{e}^{-\mathrm{i}\hat{H}_\text{e}t}\rho\,\mathrm{e}^{\mathrm{i}\hat{H}_\text{e}t}$ is then approximated by
\begin{equation}
  \mathcal{E}_\text{qDRIFT}[\rho] = \sum_{\bm{i}} p_{\bm{i}} V_{\bm{i}} \rho V_{\bm{i}}^\dagger \, ,
  \label{eq:Average_qDRIFT_Channel}
\end{equation}
where each randomized evolution sequence
\begin{equation}
  V_{\bm{i}} = \prod_{j=0}^{N-1} \mathrm{e}^{- \mathrm{i} h_{i_j} t \lambda / N} \, , \quad \text{with } \lambda = \sum_i w_i \, ,
  \label{eq:qDRIFT_Unitary}
\end{equation}
is weighted by the probability
\begin{equation}
  p_{\bm{i}} = \prod_{j=0}^{N-1} \frac{w_{i_j}}{\lambda} \, .
  \label{eq:qDRIFT_probability}
\end{equation}
In other words, each of the $N$ factors in \cref{eq:qDRIFT_Unitary} is obtained by independently drawing a term $h_i$ of \cref{eq:HamiltonianDecomposition} with probability $w_i/\lambda$ and evolving under it for the fixed time $\tau = t\lambda/N$.
Terms with large coefficients are therefore sampled more often, while the evolution time of each factor is the same regardless of the sampled term.
The channel in \cref{eq:Average_qDRIFT_Channel} approximates $\mathcal{U}_t$ with an error, in diamond norm, of $\mathcal{O}(\lambda^2 t^2/N)$~\cite{Campbell2019}.
Crucially, this bound depends on the Hamiltonian only through $\lambda$ and not on the number of terms in \cref{eq:HamiltonianDecomposition}.
This makes qDRIFT particularly appealing for electronic-structure Hamiltonians, whose number of terms grows as $\mathcal{O}(N_o^4)$, whereas the depth of Trotter circuits grows with the number of terms.

SqDRIFT uses this randomized construction to approximate the Krylov states of \cref{eq:kqd_circs}.
For each Krylov index $k$, we set $t = k\Delta t$, draw $M$ random sequences $\bm{i}$ according to \cref{eq:qDRIFT_probability}, and measure each circuit $V_{\bm{i}}\ket{\psi_0}$ in the computational basis $N_s$ times.
Since the probability of measuring the bitstring $\bm{x}$ from the state $\mathcal{E}_\text{qDRIFT}\big[\ket{\psi_0}\!\bra{\psi_0}\big]$ is
\begin{equation}
  P(\bm{x}) = \sum_{\bm{i}} p_{\bm{i}} \, \big| \bra{\bm{x}} V_{\bm{i}} \ket{\psi_0} \big|^2 \, ,
  \label{eq:qDRIFT_sampling}
\end{equation}
each sample collected in this way is distributed according to the qDRIFT approximation of the time-evolved state, without the mixed state ever being prepared explicitly.
The bitstrings collected over all Krylov indices and random realizations are then post-processed and used to build the subspace in which the Hamiltonian is diagonalized classically (see \cref{app:quantum_details} for the batching and configuration-recovery settings).
Note that, as argued in Ref.~\cite{Piccinelli2025}, it suffices that the sampling distribution, accumulated over Krylov indices and random realizations, be approximately uniform over the support of the ground state. Deviations from the exact time-evolution channel only redistribute probability among determinants, which is harmless as long as every determinant in the support retains a non-negligible sampling probability.

\subsection{Spin-adapted initial states for SqDRIFT}
\label{sec:theory-initial-states}

The choice of the initial state and of the corresponding state-preparation primitive is a key step of SqDRIFT: as discussed in Ref.~\cite{Piccinelli2025}, the ability of SqDRIFT to sample the support of the exact ground-state wavefunction efficiently depends on (and increases with) its overlap with the initial state.
When targeting a state with total spin $S$, the initial state should be an eigenstate of $\hat{S}^2$ with eigenvalue $S(S+1)$, e.g., a CSF.
In fact, it has been shown that, for strongly correlated systems, a CSF or a short linear combination of CSFs can have a substantially larger overlap with the exact ground state than a single determinant expressed in the same orbital basis~\cite{Tew2025_SpinCoupling,Ollitrault2024_OrbitalOpt}, although, for iron--sulfur clusters, optimizing the orbitals of a single determinant has been shown to yield an even larger overlap~\cite{Ollitrault2024_OrbitalOpt}.
Since the convergence guarantees of SqDRIFT become tighter as this overlap increases~\cite{Piccinelli2025}, such a choice is expected to improve the efficiency of SqDRIFT as well.
Moreover, choosing a CSF as the initial state in an SqDRIFT calculation ensures that, if the time-evolution circuit that is applied on this initial state preserves the spin (see \cref{sec:theory-circuits}), the wavefunction generated by each SqDRIFT circuit is an eigenstate of $\hat{S}^2$ as well.
Quantum circuits that prepare a CSF with a given number of electrons $N_e = N_\alpha + N_\beta$ and spin quantum numbers $S$ and $M_S$ have been proposed~\cite{Sugisaki2019_CSF-Preparation,Carbone2022_SpinEigenfunctions,Tew2025_SpinCoupling}, and can be used as the state-preparation step of SqDRIFT without modifying the rest of the workflow.
For simplicity, for given values of $N_e$ and $S$, we choose---without loss of generality---as initial state the CSF associated with the largest value $M_S = S$.
Since $\hat{H}_\text{e}$ also commutes with the spin ladder operators, $[\hat{H}_\text{e}, \hat{S}_\pm] = 0$, the $2S+1$ components $M_S = -S, \dots, S$ of a spin multiplet, which are connected by $\hat{S}_\pm$, are exactly degenerate, so that the value of $M_S$ can be chosen freely.
If the configuration has exactly $2S$ singly occupied orbitals, the CSF with $M_S = S$ corresponds to a single determinant in which all singly occupied orbitals host $\alpha$ electrons, so that $N_{\alpha} - N_{\beta} = 2S$.
Our choice is motivated by the fact that, on current noise-limited quantum processors, reducing as much as possible the depth associated with the state-preparation part enables executing deeper, qDRIFT-randomized time-evolution circuits that yield more accurate SqDRIFT calculations~\cite{Piccinelli2025}.
This choice has two further advantages.
First, the $M_S = S$ sector contains no states with total spin lower than $S$, so that its lowest root cannot acquire lower-spin character.
Second, since each factor of the spin-adapted circuits of \cref{sec:theory-circuits} commutes with $\hat{S}_\pm$ (up to the splitting error discussed there), which do not change orbital occupations, the probability of sampling a given electronic configuration from the time-evolved state does not depend on $M_S$.
Within the $M_S$ sector, this probability is distributed over up to $\binom{n_{\text{single}}}{n_{\text{single}}/2+M_S}$ determinants, with $n_{\text{single}}$ the number of singly occupied orbitals of the configuration. Since this binomial coefficient decreases with $M_S$, for a given number of samples, the sampled distribution is most concentrated and the spin-completed subspace (see \cref{sec:theory-postprocessing}) is smallest for $M_S = S$.
However, our choice restricts the initial state to configurations with exactly $2S$ singly occupied orbitals.
For target states dominated by configurations with more than $2S$ singly occupied orbitals, such as open-shell singlets, every CSF of the dominant configuration is a linear combination of several determinants, and even the preparation of a single such CSF requires circuits whose additional depth may be prohibitive on current noise-limited hardware~\cite{Sugisaki2019_CSF-Preparation,Carbone2022_SpinEigenfunctions,Tew2025_SpinCoupling, liu2026,burkat2025}.
As hardware noise decreases towards the fault-tolerant era, such CSFs, as well as linear combinations thereof, may become viable initial states.

\subsection{Spin-adapted circuits for SqDRIFT}
\label{sec:theory-circuits}

\cref{sec:theory-initial-states} discussed how to initialize quantum circuits using states with the correct spin.
Starting from these, we build the SqDRIFT circuits shown in \cref{eq:kqd_circs,eq:qDRIFT_Unitary}.
Given the commutation relationships in \cref{eq:hami_commutation}, the \textit{exact} time-evolution operator $\mathrm{e}^{-\mathrm{i}\hat{H}_\text{e}t}$ preserves the spin of the initial state $\ket{\psi_0}$.
However, product formulas, such as Trotter--Suzuki decompositions, preserve spin only if each individual term of the Hamiltonian decomposition commutes with $\hat{S}^2$. This is not the case for the spin-orbital terms of \cref{eq:mol_hami}. The same holds for qDRIFT, since each randomized realization is a product of exponentials of individual terms we refer to as \emph{excitations}.
To avoid breaking the spin symmetry in our circuit, we must rewrite the Hamiltonian in such a way that each sampled term preserves spin by itself.
To do so, we take advantage of the global SU(2) spin invariance of the non-relativistic Hamiltonian in \cref{eq:mol_hami}.
We introduce the spin-adapted excitation operators
\begin{equation}
    \label{eq:Epq}
    \hat{E}_{pq} = \hat{a}^\dagger_{p\alpha}\hat{a}_{q\alpha} + \hat{a}^\dagger_{p\beta}\hat{a}_{q\beta} \, ,
\end{equation}
which are the spin-free one-body operators and generate the subgroup U($N_o$) of spin-independent orbital rotations.
These operators are scalars with respect to spin rotations (singlets) and therefore commute with all total-spin operators ($\hat{S}^2$, $\hat{S}_z$, $\hat{S}_\pm$).
Consequently, any polynomial in the $\hat{E}_{pq}$, and the exponential of any Hermitian combination of them, also commutes with $\hat{S}^2$.
Expressed in terms of the $\hat{E}_{pq}$, the Hamiltonian reads
\begin{equation}
\label{eq:spin_pres_hami}
 \begin{aligned}
    \hat{H}_\text{e} &= \sum_{pq} h_{pq} \hat{E}_{pq}
      + \frac{1}{2}\sum_{pqrs} g_{pqrs} \left(\hat{E}_{pq}\hat{E}_{rs} - \delta_{qr}\hat{E}_{ps}\right) \\
    &= \sum_{pq} \tilde{h}_{pq} \hat{E}_{pq}
      + \frac{1}{2}\sum_{pqrs} g_{pqrs} \, \hat{E}_{pq}\hat{E}_{rs} \, ,
 \end{aligned}
\end{equation}
where $\tilde{h}_{pq} = h_{pq} - \tfrac{1}{2}\sum_{r} g_{prrq}$ collects the one-body contribution arising from the reordering of the two-body term.

The individual terms in \cref{eq:spin_pres_hami} are not Hermitian, since $\hat{E}_{pq}^\dagger = \hat{E}_{qp}$, and their exponentials are therefore not unitary.
Since $\hat{H}_\text{e}$ is Hermitian and, for real orbitals, all integrals are real, we can write $\hat{H}_\text{e} = \tfrac{1}{2}\big(\hat{H}_\text{e} + \hat{H}_\text{e}^\dagger\big)$, which yields
\begin{equation}
    \label{eq:spin_pres_hami_herm}
    \hat{H}_\text{e} = \frac{1}{2}\sum_{pq} \tilde{h}_{pq} \, \hat{T}_{pq}
      + \frac{1}{4}\sum_{pqrs} g_{pqrs} \, \hat{T}_{pqrs} \, ,
\end{equation}
with
\begin{equation}
    \label{eq:herm_terms}
    \hat{T}_{pq} = \hat{E}_{pq} + \hat{E}_{qp} \, , \qquad
    \hat{T}_{pqrs} = \hat{E}_{pq}\hat{E}_{rs} + \hat{E}_{sr}\hat{E}_{qp} \, .
\end{equation}
Each $\hat{T}_{pq}$ and $\hat{T}_{pqrs}$ is Hermitian and commutes with $\hat{S}^2$.
Conjugated pairs that coincide because of the permutational symmetry of the integrals, namely $\hat{T}_{pq} = \hat{T}_{qp}$ and $\hat{T}_{pqrs} = \hat{T}_{srqp}$ (with $g_{pqrs} = g_{srqp}$), are merged into a single term.

Finally, we take these Hermitian terms as the decomposition in \cref{eq:HamiltonianDecomposition}.
For a term with coefficient $\gamma_i$ and operator $\hat{T}_i$, we set $w_i = |\gamma_i| \, \lVert \hat{T}_i \rVert$ and $h_i = \operatorname{sgn}(\gamma_i) \, \hat{T}_i / \lVert \hat{T}_i \rVert$, which satisfies the normalization requirements of \cref{sec:theory-sqdrift}. The component of $\hat{T}_i$ proportional to the identity, which only contributes a global phase, is dropped before normalizing.
Since every $h_i$ commutes with $\hat{S}^2$, each randomized realization $V_{\bm{i}}$ in \cref{eq:qDRIFT_Unitary} would preserve the total spin of $\ket{\psi_0}$ if each factor $\mathrm{e}^{-\mathrm{i}h_i\tau}$ were implemented exactly. Under the Jordan--Wigner mapping, however, the Pauli strings of a two-body term do not, in general, all commute. We therefore partition them into $K$ mutually commuting groups, $h_i = \sum_{k=1}^{K} h_i^{(k)}$, and implement each factor as $\prod_{k=1}^{K}\mathrm{e}^{-\mathrm{i}h_i^{(k)}\tau}$, in which every group exponential is exact. Each group conserves $N_\alpha$ and $N_\beta$, so particle number and $M_S$ are preserved exactly, whereas the individual groups do not commute with $\hat{S}^2$, so the total spin is preserved only up to a splitting error of order $\tau^2$ per factor. One-body terms and two-body terms whose Pauli strings all commute ($K=1$) are implemented exactly.
We refer to the circuits built with spin-adapted excitation operators as \emph{spin-adapted}.

\subsection{Spin-adapted post-processing of bitstrings}
\label{sec:theory-postprocessing}

When the initial state is prepared in the target spin sector and the qDRIFT terms are sampled from the \textit{Hermitized} spin-adapted Hamiltonian of \cref{eq:spin_pres_hami_herm}, the circuits prepare states that are spin pure up to the splitting error of \cref{sec:theory-circuits}, in the absence of hardware noise.
However, the diagonalization subspace is built from measurement samples of these circuits.
In the SqDRIFT setting, samples from the quantum device correspond to Slater determinants, which are eigenstates of $\hat{S}_z$ but, in general, not of $\hat{S}^2$.
More importantly, the sampled subspace is not guaranteed to be closed under $\hat{S}^2$.
We say that a subspace $\mathcal{V}$ is closed under $\hat{S}^2$ if $\hat{S}^2\ket{\psi} \in \mathcal{V}$ for all $\ket{\psi} \in \mathcal{V}$.

In previous sample-based quantum algorithms, a strategy to mitigate the spin contamination of the final wavefunction has been proposed~\cite{RobledoMoreno2025}.
Let us denote by $\chi$ the set of unique bitstrings sampled from the quantum device, and by $D$ the subspace spanned by the corresponding Slater determinants.
Moreover, we denote by $\chi_\alpha$ and $\chi_\beta$ the sets of unique $\alpha$- and $\beta$-strings, respectively.
From these sets we define the subspace 
\begin{equation}
    \label{eq:D_cart}
       D_{\text{cart}} = \text{span}\left\{ \ket{\bm{a},\bm{b}} \,\big|\, \bm{a} \in \chi_{\alpha},\, \bm{b} \in \chi_{\beta} \right\} \, ,
\end{equation}
namely the subspace spanned by all determinants obtained by combining any sampled $\alpha$-string with any sampled $\beta$-string.
We will refer to this as the \textit{symmetrized cartesian subspace} produced by the collected samples.
When the numbers of $\alpha$ and $\beta$ electrons coincide, we can symmetrize the bitstrings found in the two sectors by defining
\begin{equation}
    \label{eq:sym_chi}
    \chi_{\alpha \beta} = \chi_\alpha \cup \chi_\beta 
\end{equation}
and perform the diagonalization in the subspace determined by every ordered pair that can be obtained from $\chi_{\alpha \beta}$, namely
\begin{equation}
    \label{eq:D_cart_symm}
       D^{\text{sym}}_{\text{cart}} = \text{span}\left\{ \ket{\bm{a},\bm{b}} \,\big|\, \bm{a},\bm{b} \in \chi_{\alpha\beta} \right\} \, .
\end{equation}
$D_{\text{cart}}$ extends the original subspace and mitigates spin contamination, while $D^{\text{sym}}_{\text{cart}}$ additionally enforces invariance under the global exchange of $\alpha$ and $\beta$ strings, further reducing it.
$D^{\text{sym}}_{\text{cart}}$ is the \textit{symmetrized cartesian subspace} spanned by the collected samples.
However, neither subspace is closed under $\hat{S}^2$.
This can be shown by looking at the action of the spin operator on the Slater determinants.
In atomic units ($\hbar = 1$) and in terms of $\hat{S}_\pm$, the total-spin operator reads
\begin{equation}
    \hat{S}^2 = \hat{S}_x^2 + \hat{S}_y^2 + \hat{S}_z^2 =  \hat{S}_- \hat{S}_+ +  \hat{S}_z^2 + \hat{S}_z \, ,
\end{equation}
where $\hat{S}_- \hat{S}_+$ is the only term that does not have a diagonal action on the Slater determinants.
In second quantization,
\begin{equation}
    \label{eq:spin_second_quant}
  \hat{S}_-\hat{S}_+ = \sum_{pq}
    \hat{a}^\dagger_{q\beta}\hat{a}_{q\alpha}  \hat{a}^\dagger_{p\alpha}\hat{a}_{p\beta} ,
\end{equation}
which moves an $\alpha$ electron from orbital $q$ to orbital $p$ while moving a $\beta$ electron from $p$ to $q$.
Given a Slater determinant $\ket{\bm{a},\bm{b}}$, at fixed $p \neq q$ we have
\begin{equation}
    \hat{a}^\dagger_{q\beta}\hat{a}_{q\alpha}  \hat{a}^\dagger_{p\alpha}\hat{a}_{p\beta} \ket{\bm{a},\bm{b}} = \ket{\bm{a}+\bm{e}_p - \bm{e}_q,\bm{b} - \bm{e}_p + \bm{e}_q}\,,
\end{equation}
where $\bm{e}_p$ and $\bm{e}_q$ are the $p$-th and $q$-th canonical unit vectors. This relation holds up to a fermionic sign and only if $a_q = b_p = 1$ and $a_p = b_q = 0$; otherwise, the result vanishes.
Having $\bm{a}, \bm{b} \in \chi_{\alpha \beta}$ does not imply $\bm{a}+\bm{e}_p - \bm{e}_q,\bm{b} - \bm{e}_p + \bm{e}_q \in \chi_{\alpha \beta}$, showing that $D^{\text{sym}}_{\text{cart}}$ is not closed under $\hat{S}^2$.
Of course, as the size of the sampled subspace grows it becomes increasingly likely that the strings connected by these spin-exchange moves are also included in $\chi_{\alpha\beta}$, eventually yielding a spin-closed subspace.
However, in order to reduce the (classical) computational resources required by sample-based quantum algorithms, this property is desirable at every subspace size, and for both open- and closed-shell systems.

For this reason, we adopt a different procedure to complete the set of sampled determinants such that the resulting subspace is closed under $\hat{S}^2$.
First, we note that CSFs have exactly the property that we need: every CSF is an $\hat{S}^2$ eigenstate, and therefore any subspace spanned by CSFs is closed under $\hat{S}^2$.
Moreover, CSFs are simultaneous eigenstates of $\hat{S}_z$ and $\hat{S}^2$.
This implies that the Slater determinants forming a given CSF all share the same eigenvalue $M_S$ of $\hat{S}_z$ and, since spin operators do not change orbital occupations, belong to the same electronic configuration.
In general, this does not identify a specific CSF, since the same set of Slater determinants enters several CSFs with different coefficients~\cite{Pauncz1979,Helgaker2000}.
However, we are only interested in the determinant subspace spanned by the CSFs, not in the individual CSFs, as the diagonalization is performed in the determinant basis.
This means that we only have to identify all the other Slater determinants sharing the same configuration and $M_S$, i.e., all ways of distributing the open-shell electrons among the $n_{\text{single}}$ singly occupied orbitals at fixed $M_S$, while doubly occupied and empty orbitals remain fixed.
The $n_{\text{single}}$ singly occupied orbitals host $n_{\text{single}}$ electrons, of which $n_\alpha = n_{\text{single}}/2 + M_S$ are $\alpha$ and $n_{\text{single}}/2 - M_S$ are $\beta$.
Therefore, the number of such determinants equals the number of ways of distributing the open-shell $\alpha$ electrons among the singly occupied orbitals, i.e.
\begin{equation}
    \binom{n_{\text{single}}}{n_{\alpha}} = \binom{n_{\text{single}}}{n_{\text{single}}/2 + M_S}\,,
\end{equation}
which remains small for moderate numbers of open shells (e.g., 924 for $n_{\text{single}} = 12$ and $M_S = 0$).

With this, we end up with a set of spin-completed bitstrings $\chi_{\text{spin}}$ spanning a subspace $D_{\text{spin}}$ that is closed under $\hat{S}^2$, in which the classical diagonalization is performed.
The resulting procedure, which we refer to as \emph{spin completion}, is summarized in \cref{alg:spin_completion}.

More information about how we performed the diagonalizations in the spin-adapted subspace and how we post-processed the samples and the wavefunctions can be found in \cref{app:subspacediag,app:spin-trim}.
In practice, spin completion is applied to each sub-sampled batch, after configuration recovery, and the subspace is truncated by discarding whole configurations, which preserves closure under $\hat{S}^2$.
In the next section, we present results on different chemical phenomena that we investigated using the spin-adapted workflow both in numerical simulation and on quantum hardware.
\begin{algorithm}
\caption{Spin completion of the sampled bitstrings}\label{alg:spin_completion}
\KwIn{Set $\chi$ of unique bitstrings $\bm{x}=(\bm{a},\bm{b})$ sampled from the quantum device; $(N_\alpha,N_\beta)$, $M_S=(N_\alpha-N_\beta)/2$}
\KwOut{Spin-completed set $\chi_\text{spin}$}
$\chi \leftarrow \{\bm{x}\in\chi : |\bm{a}|\!=\!N_\alpha,|\bm{b}|\!=\!N_\beta\}$\Comment*{post-select}
$\chi_\text{spin}\leftarrow\varnothing$, \ $\mathcal{C}\leftarrow\varnothing$\Comment*{done: $\mathcal{C}$}
\ForEach{$\bm{x}=(\bm{a},\bm{b})\in\chi$}{
  $\bm{n}\leftarrow\bm{a}+\bm{b}$\Comment*{configuration}
  \If{$\bm{n}\notin\mathcal{C}$}{
    $\mathcal{C}\leftarrow\mathcal{C}\cup\{\bm{n}\}$\;
    $\mathcal{O}\leftarrow\{p : n_p=1\}$, \ $n_\text{single}\leftarrow|\mathcal{O}|$\;
    $n_\alpha^\text{open}\leftarrow n_\text{single}/2+M_S$\;
    \ForEach{$\mathcal{A}\subseteq\mathcal{O}$ with $|\mathcal{A}|=n_\alpha^\text{open}$}{
      $a'_p\leftarrow 1$ if $n_p=2$ or $p\in\mathcal{A}$, else $0$\;
      $b'_p\leftarrow 1$ if $n_p=2$ or $p\in\mathcal{O}\setminus\mathcal{A}$, else $0$\;
      $\chi_\text{spin}\leftarrow\chi_\text{spin}\cup\{(\bm{a}',\bm{b}')\}$\;
    }
  }
}
\Return{$\chi_\text{spin}$}\;
\end{algorithm}

\section{Results}
\label{sec:results}

Accounting for spin symmetry is important in the description of many quantum chemical phenomena, from ground-state to excited-state properties.
In the following, we present different use cases, including experiments on real hardware.

\subsection{Excited-state characterization of 4-aminobenzonitrile (ABN)}
\label{sec:results-abn}

The study of the low-lying energy spectrum of molecular compounds is essential to characterize their photochemical properties.
In many systems, the interplay between locally excited (LE) and charge-transfer (CT) states is of particular relevance, for instance in organic light-emitting diodes based on thermally activated delayed fluorescence~\cite{Uoyama2012,Etherington2016} and in organic photovoltaics~\cite{Vandewal2014}.
An LE state occurs when an electron is promoted from an occupied to an unoccupied molecular orbital within the same localized region, or fragment, of a molecule.
These states are usually characterized by a large oscillator strength, leading to bright, fast radiative decay.
In CT states, instead, an electron is transferred from an electron-donor fragment to an electron-acceptor fragment, which can be part of the same molecule (intramolecular CT) or belong to two distinct molecules (intermolecular CT).
CT states exhibit large changes in the dipole moment, large Stokes shifts (i.e., large energy differences between the absorption and emission maxima), and lower radiative rates than pure LE states.
\begin{figure*}
    \centering
    \includegraphics{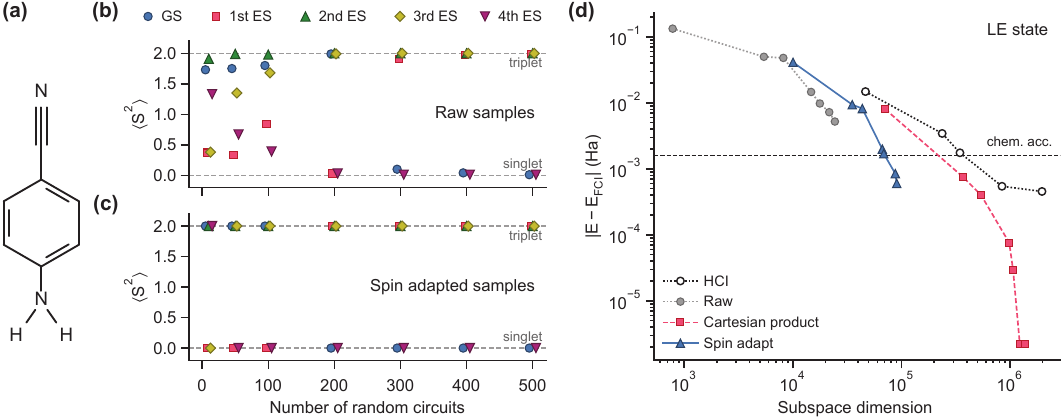}
    \caption{\textbf{Low-lying spectrum of ABN (simulation).}
      \textbf{(a)} Structure of ABN (\ce{H2N-C6H4-CN}); calculations use the 6-31G basis and an (18e, 14o) active space.
      \textbf{(b,c)} $\langle \hat{S}^2 \rangle$ of the five lowest $M_S = 0$ roots against the number of random circuits, for the raw (b) and spin-completed (c) samples; dashed lines mark the exact singlet and triplet values.
      \textbf{(d)} Energy error $|E - E_\mathrm{FCI}|$ of the LE state against the number of determinants in the subspace, for $N = 100$. Raw SqDRIFT samples, their spin completion, and the Cartesian product $D^{\text{sym}}_{\text{cart}}$ are compared against variational HCI. The dashed line marks chemical accuracy (\qty{1.6}{\milli\hartree}).}
    \label{fig:ABN_results}
\end{figure*}
Aminobenzonitriles are a widely studied class of molecules, as they are prototypical systems for the interplay between LE and intramolecular CT states~\cite{Grabowski2003,Segado2016,Loos2021}.
Their emission properties can be controlled by structural modifications or by external parameters such as solvent polarity and temperature, with important applications in materials design.
Here, we focus on 4-aminobenzonitrile (ABN, \ce{H2N{-}C6H4{-}CN}) to characterize its lowest LE and CT states.
The molecule consists of a benzene ring substituted with an electron-donating amino group (\ce{{-}NH2}) and, in the \textit{para} position, an electron-withdrawing cyano group (\ce{{-}C#N}), as shown in panel (a) of \cref{fig:ABN_results}.

From a Hartree--Fock calculation in the 6-31G basis set, we select an active space of 18 electrons in 14 orbitals, (18e, 14o).
We then emulate SqDRIFT classically, sampling the circuits with a tensor-network simulator.
For each of the three Krylov times, we draw $M = 500$ random realizations with $N = 100$ excitations each, and collect $N_s = \num{1000}$ samples per circuit.
The Hamiltonian is then diagonalized in the $M_S = 0$ subspace spanned by the unique sampled determinants.
Since we target singlet excited states, we compute $\langle \hat{S}^2 \rangle$ for the five lowest roots in order to identify and exclude triplet (and higher-multiplicity) states.
As shown in panel (b) of \cref{fig:ABN_results}, diagonalizing in the subspace spanned by the raw samples yields spin-contaminated roots when up to 100 random circuits are used, which makes it difficult to classify the excited states by their spin.
After the spin completion described in \cref{sec:theory-postprocessing}, every root is a spin eigenfunction, here a singlet or a triplet, at every number of circuits (panel (c)).

Although this is not a general criterion for distinguishing LE from CT states, we use the total spin and the change in dipole moment along the main molecular axis to identify the first singlet excited state as the LE state.
Panel (d) shows the convergence of its energy towards the FCI reference as a function of the dimension of the diagonalized matrix for four subspace constructions: the raw SqDRIFT samples, their spin completion, the Cartesian product $D^{\text{sym}}_{\text{cart}}$ of \cref{eq:D_cart_symm}, and, as a classical selected-CI reference, variational heat-bath configuration interaction (HCI)~\cite{Holmes2016,Sharma2017,Holmes2017_excited}, as implemented in DICE~\cite{Dice}.
The raw samples converge slowly and remain above chemical accuracy over the whole range, with an error of about \qty{5}{\milli\hartree} at the largest subspace of about \num{2.5e4} determinants.
Spin completion reaches chemical accuracy at about \num{7.5e4} determinants, and an error of about \qty{0.6}{\milli\hartree} below \num{1e5} determinants.
The Cartesian product eventually attains the smallest errors, down to about \qty{2e-6}{\hartree}, but only at about \num{1.4e6} determinants, roughly \qty{35}{\percent} of the FCI space (\num{4.0e6} determinants), and it reaches chemical accuracy only beyond about \num{2.5e5} determinants.
HCI reaches chemical accuracy at about \num{5e5} determinants and an error of about \qty{0.5}{\milli\hartree} at its largest subspace of \num{2e6} determinants.
The large subspaces produced by HCI are mainly due to the fact that we are considering the union of all the subspaces relevant to the excited states up to the target LE one~\cite{Holmes2017_excited}.
Among the constructions considered, spin completion therefore provides the most compact subspace at chemical accuracy, about three and seven times smaller than those required by the Cartesian product and HCI, respectively.
More details on the calculations on ABN are given in \cref{app:classical_details}.

\subsection{Spin-state splitting in an \texorpdfstring{\ce{Fe^{II}N4}}{Fe(II)N4} model complex}
\label{sec:results-fen4}

\begin{figure*}
    \centering
    \includegraphics{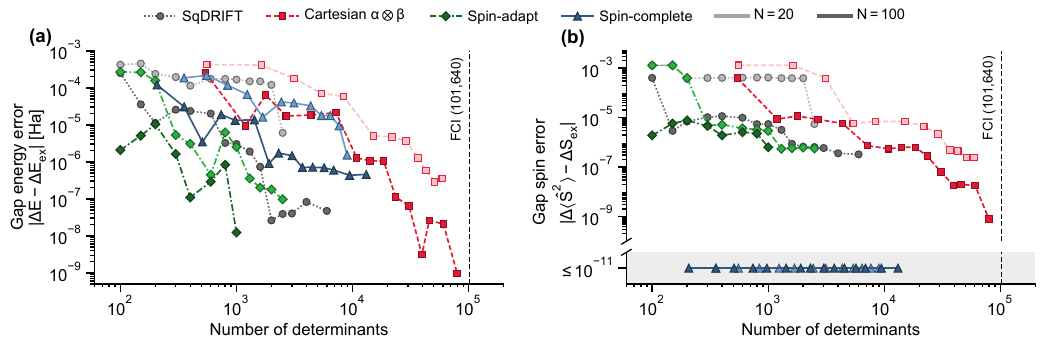}
    \caption{\textbf{Spin-state splitting in an \ce{FeN4} centre (simulation).} Error on the quintet--triplet gap
    $\Delta = E_{S=2}-E_{S=1}$, with the two states computed as the lowest roots of the $M_S = 2$ and $M_S = 1$ sectors and in an (8e, 11o) active space, against the total number of determinants in the two subspaces. \textbf{(a)} Gap energy error $|\Delta E - \Delta E_{\mathrm{ex}}|$ and \textbf{(b)} gap spin error $|\Delta \langle \hat{S}^2 \rangle - \Delta S_{\mathrm{ex}}|$ (with $\Delta S_{\mathrm{ex}}\equiv\Delta S(S+1)$), for the raw sampled determinants (SqDRIFT), their Cartesian product $\alpha \otimes \beta$, the spin-completed and target-spin-projected subspace (spin-complete), and spin-adapted sampling. The latter is the raw-determinant construction applied to counts from the spin-adapted circuits, without completion or projection: although the circuits prepare states that are spin pure up to the splitting error of \cref{sec:theory-circuits}, the sampled set of determinants is not closed under $\hat{S}^2$, which causes the residual $\langle \hat{S}^2 \rangle$ error. Marker shape denotes the construction and color the number of excitations ($N = 20$, $100$); the dashed line marks the summed full CI dimension of the two sectors (\num{101640}). Spin errors of the spin-complete construction are below \num{e-11}, i.e., at solver precision, and are drawn clamped onto the shaded band.}
    \label{fig:FeN4}
\end{figure*}

Iron--nitrogen sites are the catalytic motif of \ce{Fe{-}N{-}C} materials for the oxygen reduction reaction~\cite{Lefevre2009,Zitolo2015}, whose activity depends on the spin state of the iron centre~\cite{FeNC_NatCommun2022,FeNC_JMCA2023}. 
For the parent iron(II) porphyrin, the ordering of the triplet and quintet states is a notoriously difficult multi-reference problem, which requires large active spaces and self-consistent orbital optimization to be resolved~\cite{LiManni2016_FeP,LiManni2018_FeP,OlivaresAmaya2015}. 
The quantity of interest is therefore not a single total energy but a \emph{difference} between states of different multiplicity, which places a correspondingly sharper demand on the subspace: the two legs of the splitting must be converged consistently, and each must be an eigenstate of $\hat{S}^2$ for the difference to correspond to a physical spin-state gap at all.

We examine the quintet--triplet splitting $\Delta = E_{S=2}-E_{S=1}$ of a \ce{[Fe(NH3)4]^{2+}} (in short \ce{FeN4}) complex using an (8e, 11o) active space, where the FCI splitting is \qty{0.3048}{\hartree} (\qty{191}{\kcal\per\mole}) and the exact spin difference is $\Delta S(S+1) = 4$.
The SqDRIFT circuits start from a single determinant in each sector: the Hartree--Fock determinant for $M_S = 1$ ($\alpha$ orbitals $\{0,\dots,4\}$, $\beta$ orbitals $\{0,1,2\}$) and, for $M_S = 2$, the determinant minimizing $\langle D | \hat{H}_\text{e} | D \rangle$ ($\alpha$ orbitals $\{0,1,2,3,4,9\}$, $\beta$ orbitals $\{0,3\}$), which lies \qty{0.109}{\hartree} below the Hartree--Fock determinant in this orbital basis.
In both determinants all open-shell electrons have $\alpha$ spin, so each is an exact eigenfunction of $\hat{S}^2$ with $S = M_S$, and the circuits start in the target spin sector.
The $M_S = 2$ and $M_S = 1$ sectors are sampled independently and paired at equal numbers of sampled bitstrings---the sampler's own control variable---since the same number of samples expands to different determinant counts in each sector.

\cref{fig:FeN4} shows that only the spin-complete construction yields a gap between genuine spin eigenstates: its spin error is zero to solver precision at every subspace size, and it reaches a gap energy error of \qty{5.6e-9}{\hartree} at a total dimension of \num{1.1e4}, an order of magnitude below the full CI dimension of the pair. The raw and spin-adapted subspaces carry residual spin errors of \numrange{1e-7}{10e-7}, comparable to their energy errors, so their apparent accuracy on $\Delta$ is not accompanied by spin purity. The Cartesian product achieves the lowest absolute error (\qty{1.3e-10}{\hartree}) only by diagonalizing \num{9.1e4} determinants, within \qty{10}{\percent} of the full space.

Because the error cancellation between the two legs is accidental, a gap can appear more accurate than either sector individually, which explains the non-monotonic curves. Moreover, the cancellation is strongest for the spin measure, where two equally contaminated legs can produce a small difference in $\langle \hat{S}^2 \rangle$ while neither state is spin pure.

\subsection{Spin energetics of the \texorpdfstring{\oxoMn{}}{oxo-Mn(Salen)} complex}
\label{sec:results-oxo}
\begin{figure*}
    \centering
    \includegraphics{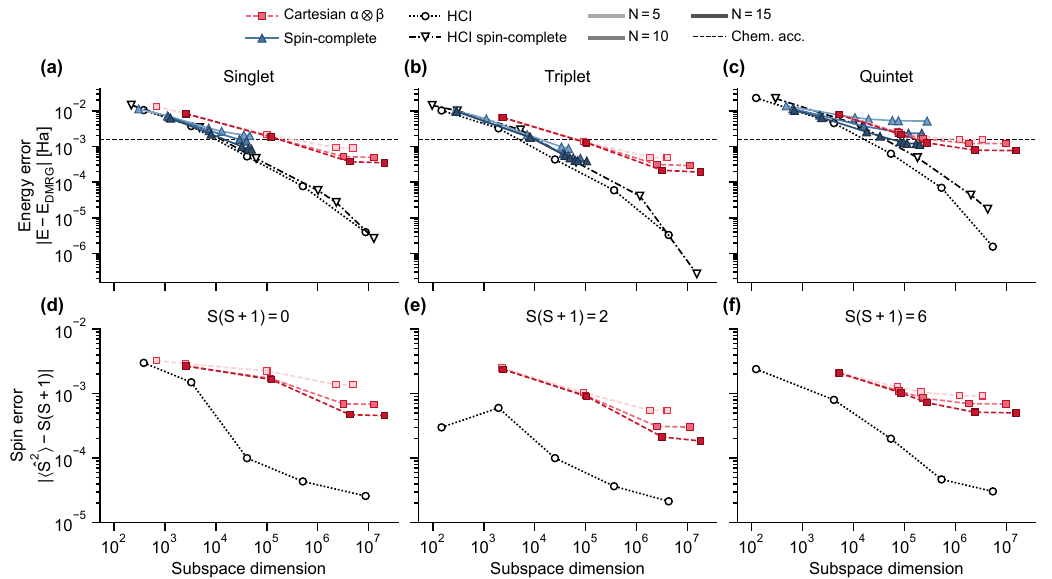}
    \caption{\textbf{Spin energetics of the \oxoMn{} molecular compound (hardware).} Energy error $|E - E_{\mathrm{DMRG}}|$ \textbf{(a--c)} and spin error
    $|\langle \hat{S}^2 \rangle - S(S+1)|$ \textbf{(d--f)} against the subspace dimension, for the singlet, triplet, and quintet states, computed in the $M_S = 0$, 1, and 2 sectors, respectively, in the (32e, 24o) active space. Marker shape denotes the subspace construction and color the number of excitations ($N = \{5, 10, 15\}$); HCI and its spin-adapted variant (see main text and \cref{app:classical_details}) are shown as references in red. 
    The dashed line marks chemical accuracy (\qty{1.6}{\milli\hartree}). 
    Classical and quantum spin-complete errors are omitted from the plot, as their values are substantially lower than those of the other methods (machine precision).
    }
    \label{fig:oxo_results}
\end{figure*}
High-valent manganese--oxo species are the proposed active intermediates of the Jacobsen--Katsuki epoxidation~\cite{Zhang1990,Irie1990,McGarrigle2005}, and their strongly correlated Mn~$3d$/O~$2p$ manifold makes them a standard benchmark for multireference electronic-structure methods. The \oxoMn{} complex studied here is small enough to admit a converged DMRG reference in a (32e, 24o) active space (see \cref{app:classical_details}), yet retains the dense low-lying spin spectrum that makes spin symmetry difficult to control.

Previous studies indicate that the ground state of \oxoMn{} is a singlet, closely followed by a triplet, whereas the lowest quintet lies significantly higher in energy~\cite{Ivanic2004,Sears2006,Wouters2014}. 
Our DMRG references in the (32e, 24o) active space reproduce this ordering: the triplet lies only \qty{0.48}{\milli\hartree} (\qty{0.30}{\kcal\per\mole}) above the singlet, while the quintet is higher by \qty{149}{\milli\hartree} (\qty{94}{\kcal\per\mole}). 
Each state is computed as the lowest root of its own $M_S$ sector ($M_S = 0$, 1, and 2 for the singlet, triplet, and quintet, respectively).
The triplet and quintet legs are thus protected from lower-spin contamination by construction, whereas the $M_S = 0$ sector contains both the singlet and the $M_S = 0$ component of the triplet, only \qty{0.48}{\milli\hartree} apart.
This near-degeneracy makes the singlet leg a stringent test of spin purity: the residual energy errors of the sampled subspaces discussed below are of the same order as the gap, so in a subspace that is not closed under $\hat{S}^2$ nothing prevents the lowest $M_S = 0$ root from acquiring triplet character, and its multiplicity can only be checked \textit{a posteriori} through $\langle \hat{S}^2 \rangle$.

\cref{fig:oxo_results} compares the two SqDRIFT constructions against HCI for each multiplicity. The spin errors (panels d--f) separate the methods unambiguously. The spin-completed subspaces are spin pure to solver precision at every subspace size and for all three states, so their curves sit on the clamped band
throughout. The Cartesian product, by contrast, retains a contamination of \numrange{4.6e-4}{9.3e-4} in $\langle \hat{S}^2 \rangle$ even at its largest subspaces, decreasing only slowly with $N$; the Cartesian product is not closed under $\hat{S}^2$, and spin purity is recovered only incidentally as the subspace approaches the full space. HCI was performed in two different ways: we first ran DICE with different variational thresholds (HCI line in the figure), then we applied the spin-completing post-processing described in \cref{sec:theory-postprocessing} and performed a final subspace diagonalization (HCI spin-adapt curve).

The energy errors (panels a--c) show why this matters. 
At $N = 15$ the spin-complete subspace reaches \qty{0.84}{\milli\hartree} for the singlet---within chemical accuracy---at a diagonalized dimension of \num{5.0e4}, whereas the Cartesian product attains a lower \qty{0.35}{\milli\hartree} only at \num{2.0e7} determinants, a subspace larger by a factor of \num{\sim 400}. 
The same pattern holds for the triplet (\qty{0.39}{\milli\hartree} at \num{1.1e5} versus \qty{0.19}{\milli\hartree} at \num{1.8e7}) and for the quintet. 
Therefore, even on hardware and in the presence of noise, spin adaptation buys exact spin symmetry \emph{and} a reduction in the matrix to be diagonalized, at comparable accuracy (both within chemical accuracy) and at fixed $N$.

HCI remains more accurate in absolute terms, reaching errors of the order of \qty{e-6}{\hartree}, but at subspace dimensions of \numrange{4e6}{15e6} and without exact spin purity.
For this system, the residual contamination of the Cartesian-product and HCI roots is small in practice.
As a rough estimate, a deviation $\delta$ in $\langle \hat{S}^2 \rangle$ corresponds to a weight of $\delta/2$ ($\delta/6$) of the triplet (quintet) in the singlet root, and of $\delta/4$ of the quintet in the triplet root; for $\delta \lesssim \num{e-3}$, this shifts the energies by at most about \qty{0.03}{\milli\hartree}, well below the singlet--triplet gap.
The advantage of spin adaptation is therefore not the repair of a contamination large enough to swap the two states, but a guarantee: in a spin-complete subspace every root is an exact spin eigenfunction, so the multiplicity of the lowest $M_S = 0$ root is determined exactly rather than inferred from $\langle \hat{S}^2 \rangle$.
This guarantee comes with a diagonalized dimension two to three orders of magnitude smaller than that of the Cartesian product at the same number of excitations $N$. 
For more details on the chemical calculations and the quantum experiments, see \cref{app:classical_details,app:quantum_details}.

\section{Conclusions}
\label{sec:conclusions}

In this work, we have introduced a general spin-adapted formulation of SqDRIFT.
We have discussed how its three key components---the initial-state preparation, the time-evolution circuits, and the classical post-processing---can be generalized to account for spin symmetry.
Based on both classical simulations and experiments on quantum hardware, we have demonstrated that the resulting spin-adapted SqDRIFT variant significantly reduces the classical overhead, i.e. the dimension of the diagonalized subspace, while guaranteeing spin-pure roots.

Our hardware experiments employed only the spin-adapted classical post-processing, i.e., spin completion (\cref{sec:theory-postprocessing}); future work will also execute spin-adapted circuits on quantum hardware.
This could require future generations of quantum devices, since preparing generic CSFs and implementing spin-adapted SqDRIFT circuits (for a fixed number of excitations $N$) substantially increase the circuit depth. As an example, for the \ce{FeN4} model, the two-qubit depth before routing is 8.8 and 4.3 times that of the standard circuits for $N=20$ and $N=100$, respectively.

Although our framework relies on the Jordan--Wigner fermion-to-qubit mapping, the space spanned by the CSFs could also be encoded directly in the qubit register, using, e.g., the mappings of Refs.~\cite{burkat2025,Gandon2025_SpinAdapted}.
On the one hand, such mappings would make the adaptation of the post-processing unnecessary, since each sample would directly correspond to a CSF.
On the other hand, the time-evolution operator would become considerably deeper in this basis, so that such approaches would be practical only on fault-tolerant hardware.

Finally, although we have focused on SqDRIFT, the framework readily extends to SKQD~\cite{Yu2025}.
Spin completion also applies directly to SQD variants based on variational ans\"atze, whereas enforcing spin symmetry at the level of the variational circuit is less straightforward: SQD has mostly been combined with the LUCJ ansatz~\cite{Motta2023_LUCJ}, which does not conserve the total spin.
However, spin-adapted ans\"atze tailored to molecular wave functions~\cite{Kjellgren2026_OO-VQE}, including coupled-cluster-based ones~\cite{Mehendale2023_SpinAdapted-UCC}, have recently been proposed and could be combined with the framework presented here.

\section{AI use statement}
\label{sec:ai_use}
We used large language models to assist with writing and debugging parts of the code and with language editing of the manuscript.
The authors designed the method and the experiments, checked the code, the results and every cited work, and take full responsibility for the content of the paper.

\section{Author contributions}
\label{sec:contributions}
S.B. conceived the idea, performed classical calculations and experiments on quantum hardware.
S.B. and S.P. wrote the code for the spin-adapted workflow.
S.P. and A.B. performed classical quantum chemistry calculations, classical simulations and experiments on quantum hardware.
A.B., F.T. and I.T. supervised the project.
All authors wrote and revised the manuscript.

\begin{acknowledgments}
We thank Davide Cugini, Davide Zanin, Laurin Fischer, Javier Robledo-Moreno and Max Rossmannek for fruitful discussions.
This research was supported by the project RESQUE (Rethinking Quantum Simulations in the Quantum Utility Era, grant number \texttt{20QU-1\_225229}) and by the National Center of Competence in Research (NCCR) SPIN (grant number 225153), both funded by the Swiss National Science Foundation.
\end{acknowledgments}

\bibliography{references}

\clearpage
\newpage

\appendix
\crefalias{section}{appendix}
\onecolumngrid

\section{Diagonalizing in the spin-adapted subspace}
\label{app:subspacediag}

\cref{sec:theory-postprocessing} showed how the sampled determinants can be post-processed so that the diagonalization subspace $D_{\text{spin}}$ is closed under $\hat{S}^2$.
The orthogonal projector $P$ onto $D_{\text{spin}}$ then commutes with $\hat{S}^2$, and so does the projected Hamiltonian, $[P\hat{H}_\text{e}P, \hat{S}^2] = P[\hat{H}_\text{e}, \hat{S}^2]P = 0$.
Its eigenfunctions can therefore be chosen with a definite total spin (and are, barring accidental degeneracies between different spin sectors), so they can be filtered directly by their spin.

When only a specific spin sector is of interest, computational effort spent on other sectors can be avoided.
Here, we illustrate an additional step that can be performed by taking advantage of the structure of the spin-adapted subspace and that can be used to guarantee that Davidson--Liu~\cite{Davidson1975,Liu1978} diagonalization is performed in a fixed-total-spin subspace.
Diagonalizing $\hat{S}^2$ yields the $n_S$ orthonormal eigenvectors with the target spin, which form the columns of the isometry $V \in \mathbb{R}^{|D_{\text{spin}}| \times n_S}$, where $|D_{\text{spin}}|$ is the dimension of the spin-completed subspace~\cite{Pauncz1979}.
Each configuration with $n_{\text{single}}$ open shells contributes $f(n_{\text{single}},S) = \binom{n_{\text{single}}}{n_{\text{single}}/2-S} - \binom{n_{\text{single}}}{n_{\text{single}}/2-S-1}$ columns, so that $n_S = \sum f(n_{\text{single}},S)$ over the configurations in $D_{\text{spin}}$.
Crucially, the spin operator is block diagonal per configuration, as it only rearranges the spins of the open shells and it does not change which orbitals are doubly or singly occupied.
At fixed $M_S$ each block depends only on $n_{\text{single}}$; its eigenvectors are therefore computed once per distinct open-shell count and reused.
In this work, we constructed the matrix of $\hat{S}^2$ explicitly; for larger systems, schemes that never form this matrix, such as GUGA~\cite{Shavitt2009_GUGA}, may become necessary.
Once the matrix $V$ is obtained, we define the transformed Hamiltonian
\begin{equation}
    \label{eq:single-spin-hami}
    G = V^{T} H V \,, \quad \text{with } G \in \mathbb{R}^{n_S \times n_S} \,,
\end{equation}
and use this to perform the Davidson--Liu diagonalization in a single spin sector.
The iterations are started from the determinant used as the initial state of the SqDRIFT circuits, projected onto the target spin sector. We refer to this restriction of the diagonalization to a single total-spin sector as \emph{spin projection}.
The Davidson method belongs to a class of algorithms that approximate the eigenvalues and eigenvectors of a large matrix by projecting it onto a smaller subspace and minimizing the Rayleigh quotient~\cite{Saad2011}.
The Rayleigh--Ritz procedure also provides an estimate of the distance from the exact eigenpair, which is used to expand the search subspace iteratively.
In particular, given ($u_p$, $\theta_p$) as the pair of approximated eigenvector and eigenvalue at iteration $p$ of the Davidson algorithm, the residual vector is:
\begin{equation}
    r_p = H u_p - \theta_p u_p \, .
\end{equation}
While the exact correction that would make $\lVert r_p \rVert$ vanish cannot be computed, Davidson proposed to approximate the optimal correction vector at every iteration as
\begin{equation}
    t_p = \frac{- r_p}{\text{diag}(H) - \theta_p I} \,,
\end{equation}
where $(\operatorname{diag}(H) - \theta_p I)^{-1}$ acts as a diagonal preconditioner (the division is element-wise).

In the projected case, the same iterations are performed for $G$ in $\mathbb{R}^{n_S}$.
Given the approximate eigenpair $(u_p, \theta_p)$, with $u_p \in \mathbb{R}^{n_S}$, the residual reads
\begin{equation}
    r_p = G u_p - \theta_p u_p = V^{T} \big( H V u_p \big) - \theta_p u_p \, ,
    \label{eq:projected_residual}
\end{equation}
so that $G$ never needs to be formed explicitly: $V u_p$ is mapped back to the determinant basis, $H$ acts there, and the result is projected again onto the target spin sector.
The correction vector is built from the diagonal of $G$,
\begin{equation}
    (t_p)_i = \frac{-(r_p)_i}{G_{ii} - \theta_p} \, , \qquad
    G_{ii} = \sum_{\mu\nu} V_{\mu i} H_{\mu\nu} V_{\nu i} \, ,
\end{equation}
where $\mu$ and $\nu$ run only over the determinants of the configuration to which column $i$ belongs, since $V$ is block diagonal.
This projection is not required to obtain spin-pure wave functions, since the eigenfunctions in $D_{\text{spin}}$ already have a definite total spin; it reduces the dimension of the diagonalized matrix and the number of roots required when a given multiplicity is targeted.
All spin-completed results in \cref{sec:results} use it, with the target multiplicity of each state.

\section{Spin-preserving post-processing of the wave function}
\label{app:spin-trim}

Noise in the quantum device broadens the distribution of the bitstrings sampled from the prepared circuits.
This affects the quantum-selected subspace by introducing determinants with negligible weight in the low-energy states (the so-called deadwood).

For both constructions, noise is addressed via the self-consistent configuration recovery of Ref.~\cite{RobledoMoreno2025}; for the spin-completed subspaces, bitstrings that still have the wrong $(N_\alpha, N_\beta)$ after recovery are discarded before spin completion. Deadwood, on the other hand, is removed \textit{a posteriori} by discarding determinants that contribute negligibly to the low-energy states.
This procedure, often called wave-function trimming, has been used in both classical and quantum subspace methods~\cite{zhang2025, merz2026} to reduce the computational effort of subsequent processing of the wave function, without compromising the accuracy of the method.

Trimming is performed by ordering the determinants by their CI contribution and removing those whose amplitudes sit below a given threshold.
The squared amplitudes are commonly used as a proxy for the contribution of each determinant to the energy.
However, this might compromise the spin purity of our wave function.
For this reason we perform a spin-preserving trimming, following Ref.~\cite{Bytautas2007}.
In this approach, the determinants are again grouped by electronic configurations and the contribution of a single configuration is computed as
\begin{equation}
    \label{eq:config_trim}
    C_{\text{ec}}^2 = \sum_{i=0}^{N_{\text{det}}-1} c^{2}_i \, ,
\end{equation}
where $N_{\text{det}}$ is the number of determinants belonging to the same electronic configuration and $c_i$ are the CI coefficients of such determinants, as defined in \cref{eq:slater_wf}.
Discarding whole configurations according to this weight yields truncated CI expansions that remain eigenfunctions of $\hat{S}^2$---provided the untrimmed wavefunction is itself an eigenfunction of $\hat{S}^2$---since $\hat{S}^2$ is block diagonal in the configurations.
In Ref.~\cite{Bytautas2007}, another criterion is discussed by introducing the mean electronic configuration contribution
\begin{equation}
    \bar{C}_{\text{ec}}^2 = \frac{\sum_{i=0}^{N_{\text{det}}-1} c^{2}_i}{N_{\text{det}}} \, ,
\end{equation}
which avoids favoring configurations whose large total weight stems from many low-weight determinants.
In this work, we truncated by configuration weight, which resulted in the most compact wavefunctions at a given accuracy.
All hardware experiments on the \oxoMn{} complex use configuration-weight truncation.

\section{Details on the classical chemical calculations}
\label{app:classical_details}

This Appendix collects the settings of the classical calculations underlying \cref{sec:results}; \cref{fig:detail-calculations-abn} and \cref{fig:detail-calculations-fen4}provide additional information on the subspace sizes and spin errors discussed there.

\paragraph*{ABN.}
The (18e, 14o) active space was selected from a Hartree--Fock calculation in the 6-31G basis set using the AVAS active space selection method~\cite{Sayfutyarova2017} (see \cref{tab:ABN-geometry} for the Cartesian coordinates of the molecule).
Let $n_\text{str}=\binom{14}{9}=\num{2002}$ be the number of spin strings with 9 electrons in 14 orbitals. The FCI reference was computed in the full $M_S = 0$ sector of the active space---which contains $D_\mathrm{FCI}=n_\text{str}^2 = \num{4008004}$ determinants---and the HCI references with DICE~\cite{Dice} for variational thresholds $\varepsilon_1 = 10^{-2} \rightarrow 10^{-6}$.
\paragraph*{\ce{FeN4} model.}
The \ce{[Fe(NH3)4]^{2+}} complex is a square-planar model with \ce{Fe-N} distance fixed at the iron porphyrin value (see \cref{tab:FeN4-geometry} for the Cartesian coordinates of the molecule).
The (8e, 11o) active space is prepared with AVAS~\cite{Sayfutyarova2017} after a Hartree--Fock calculation in the def2-SVP basis set~\cite{Weigend2005}.
The $M_S = 1$ ($N_\alpha = 5$, $N_\beta = 3$) and $M_S = 2$ ($N_\alpha = 6$, $N_\beta = 2$) sectors contain \num{76230} and \num{25410} determinants, respectively.
\paragraph*{\oxoMn{}.}
The geometry of the \oxoMn{} complex was optimized for the singlet state at the CASSCF(10e,10o) level of theory, using the 6-31G basis set~\cite{Ivanic2004,OlivaresAmaya2015}. 
The corresponding Cartesian coordinates are provided in \cref{tab:oxoMn-geometry}. 
Following the procedure described in Ref.~\cite{OlivaresAmaya2015}, we selected an active space comprising 32 electrons in 24 orbitals (32e, 24o) and performed DMRG calculations to obtain reference wavefunctions and energies.
HCI calculations for the complex were performed with DICE~\cite{Dice} without PT2 correction and with variational thresholds $\varepsilon_1 = 10^{-2} \rightarrow 10^{-6}$.
For the HCI spin-adapt calculations, we collected the samples obtained with DICE at a fixed variational threshold $\varepsilon_1$, applied the spin-completion procedure described in \cref{sec:theory-postprocessing}, and performed a final diagonalization in DICE keeping this spin-completed subspace fixed.
CSF-based variants of HCI have been proposed~\cite{Ugandi2023}, but their usage would require a completely new code, which is beyond the scope of this work.

\section{Details on the quantum experiments}
\label{app:quantum_details}

\paragraph*{ABN.}

\begin{figure}
    \centering
    \includegraphics{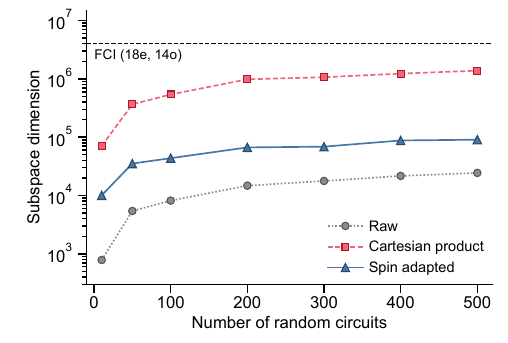}
    \caption{\textbf{Subspace sizes for ABN (simulation).}
      Dimension of the ABN subspace in the $M_S = 0$ sector as a function of the number of random circuits sampled, for $N = 100$: raw SqDRIFT samples, their spin completion, and the Cartesian product. 
      The dashed line marks the FCI dimension (\num{4008004}).
      }
    \label{fig:detail-calculations-abn}
\end{figure}

\begin{figure}
    \centering
    \includegraphics{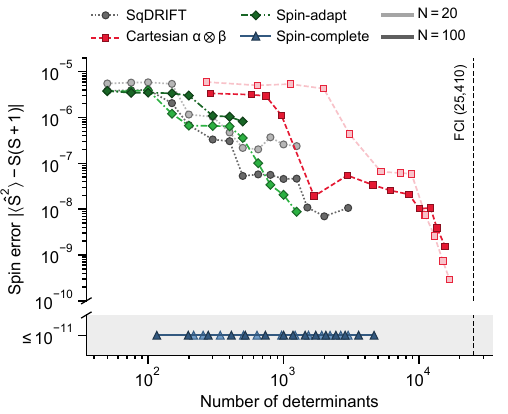}
    \caption{\textbf{Spin purity for the \ce{FeN4} quintet (simulation).}
    Spin error $|\langle \hat{S}^2 \rangle - S(S+1)|$ of the lowest $M_S = 2$ root of the \ce{FeN4} model ($S = 2$) against the number of determinants, for the four constructions of \cref{fig:FeN4}. Marker shape denotes the construction and color the number of excitations ($N = 20$, $100$; spin-adapted sampling at $N = 20$ only). The dashed line marks the FCI dimension of the sector (\num{25410}). Spin errors of the spin-complete construction are below \num{e-11}, i.e., at solver precision, and are drawn clamped onto the shaded band.}
    \label{fig:detail-calculations-fen4}
\end{figure}

To showcase the capabilities of the spin-adapted workflow in the study of excited states, we performed a classical simulation of SqDRIFT on the ABN model.
For each of the three Krylov times, we draw $M = 500$ random realizations with $N = 100$ excitations each, and collect $N_s = \num{1000}$ samples per circuit.
\cref{fig:detail-calculations-abn} shows how the size of the subspaces of \cref{fig:ABN_results}(d) grows with the number of circuits.
As explained in the main text, the dimension of the symmetrized Cartesian product is given by the product of the numbers of distinct $\alpha$ and $\beta$ strings (the square of $|\chi_{\alpha\beta}|$ given the symmetrization), each bounded by $n_\text{str}$.
The Cartesian product reaches about $n_\text{str}^2\approx 10^6$ determinants at 200 circuits and grows slowly afterwards, covering about one third of the FCI space at 500 circuits.
Spin completion enlarges the raw subspace by a factor that decreases from about twelve at 10 circuits to about four at 500, since an increasing fraction of the determinants added by completion has already been sampled.
At 500 circuits, the spin-completed subspace is still more than an order of magnitude smaller than the Cartesian product.
\paragraph*{\ce{FeN4}.}
SqDRIFT was performed using a classical noiseless simulator on the (8e,11o) active space of the \ce{FeN4} model in order to study the quintet--triplet gap.
For each sector and each of the Krylov times $t = \{1, 2, 3\}$, we sampled $M = 100$ random circuits with $N = \{20, 100\}$ excitations, collecting $N_s = 512$ samples per circuit.
Since $N_\alpha \neq N_\beta$, the Cartesian product is $D_{\text{cart}}$ of \cref{eq:D_cart}, without symmetrization.
$\langle \hat{S}^2 \rangle$ is evaluated on the eigenvector embedded in the full $M_S$ sector.
For the spin-complete construction, the $x$-axis of \cref{fig:FeN4,fig:detail-calculations-fen4} counts the determinants of $D_\text{spin}$; the matrix diagonalized after spin projection (\cref{app:subspacediag}) is smaller.

\cref{fig:detail-calculations-fen4} resolves the quintet leg of the gap in \cref{fig:FeN4}.
The $M_S = 2$ sector contains no states with $S < 2$ ($\langle \hat{S}^2 \rangle \geq 6$), and a deviation $\delta$ corresponds to a mixture of $S = 3$ and $S = 4$ components with total weight between $\delta/14$ and $\delta/6$.
The spin-complete construction is spin pure to solver precision at every size by definition.
The raw and spin-adapted-sampling subspaces show spin errors that decrease from about $10^{-5}$ at 50 determinants to below $10^{-8}$ at \num{3000}, with no appreciable difference between the two samplers.
This is consistent with \cref{sec:theory-postprocessing}: the residual error stems from the sampled set of determinants not being closed under $\hat{S}^2$, not from the circuits.
The Cartesian product reaches errors of about $10^{-10}$ only as its dimension approaches that of the full sector.
Finally, in this sector spin completion enlarges the sampled subspace by less than a factor of three, compared with the factor of four to twelve found for ABN at $M_S = 0$.
This agrees with the argument of \cref{sec:theory-initial-states} that completion adds the fewest determinants for $M_S = S$.
%
\paragraph*{\oxoMn{}.}
Experiments on the spin-state energetics of the \oxoMn{} complex were performed on the IBM Heron r3 quantum processing unit (QPU), \texttt{ibm\_boston}. 
For each spin state, we applied SqDRIFT using $k=3$ Krylov times and $M=500$ randomly generated circuit realizations for each of $N=5,10,$ and $15$ excitations. 
The quantum circuits were initialized with different states for singlet, triplet and quintet study. 
As discussed in the main text, for simplicity we considered initial states with $M_S = S$ with $2S$ singly occupied orbitals, represented by a single determinant.
Given the (32e, 24o) active space we prepared a $(16 \alpha, 16 \beta)$ initial state for the singlet, $(17 \alpha, 15 \beta)$ for the triplet and $(18\alpha, 14\beta)$ for the quintet.
When preparing the Hamiltonian for the qDRIFT randomization protocol, terms diagonal in the occupation-number basis are excluded from the decomposition, so the circuits approximate the evolution under the off-diagonal part of $\hat{H}_\text{e}$.
Given that only the sampled support enters the subspace, this does not affect the method.
We additionally tracked the sampled excitations to identify those that had no effect on the quantum state.
Such trivial excitations, which include, for example, excitations of electrons between already occupied orbitals, were discarded and resampled. 
This filtering affects only the final circuit depth, as the qDRIFT evolution time is adjusted to account for the discarded excitations. 
In this way, we avoid introducing additional noise associated with unnecessary circuit operations.
Finally, each circuit was sampled $N_s=1000$ times, yielding a total of 1.5 million shots per spin state and value of $N$, i.e., 4.5 million per spin state and 13.5 million in total. 


\begin{table}
  \centering
  \caption{Optimized Cartesian coordinates (\si{\angstrom}) of ABN.}\vspace*{5pt}
  \label{tab:ABN-geometry}
  \setlength{\tabcolsep}{2pt}
  \begin{tabular*}{0.9\textwidth}{@{\extracolsep{\fill}} c S[table-format=-1.10] S[table-format=-1.10] S[table-format=-1.10] @{\hspace{1.5em}} c S[table-format=-1.10] S[table-format=-1.10] S[table-format=-1.10]}
    \toprule
    {Atom} & {$x$} & {$y$} & {$z$} & {Atom} & {$x$} & {$y$} & {$z$} \\
    \midrule
\ce{N} &  3.22465175 & -0.00016435 & -0.00016248 & \ce{C} & -2.37901020 &  0.00002585 &  0.00007608 \\
\ce{N} & -3.52808419 &  0.00003935 &  0.00012624 & \ce{H} &  1.67124777 & -2.13635152 & -0.00008321 \\
\ce{C} &  1.85428028 &  0.00004661 & -0.00011584 & \ce{H} &  1.67125967 &  2.13648964 & -0.00011838 \\
\ce{C} &  1.13987449 & -1.20419643 & -0.00007314 & \ce{H} & -0.77189917 & -2.13075023 &  0.00002373 \\
\ce{C} &  1.13990772 &  1.20432135 & -0.00009317 & \ce{H} & -0.77189568 &  2.13073759 & -0.00001173 \\
\ce{C} & -0.94910189 &  0.00001219 &  0.00001215 & \ce{H} &  3.73679291 &  0.84648597 & -0.00032793 \\
\ce{C} & -0.23864408 & -1.20022502 & -0.00000950 & \ce{H} &  3.73662970 & -0.84690143 & -0.00031335 \\
\ce{C} & -0.23860907 &  1.20023043 & -0.00002947 &        &             &             &             \\
    \bottomrule
  \end{tabular*}
\end{table}
\begin{table}
  \centering
  \caption{Optimized Cartesian coordinates (\si{\angstrom}) of \ce{[Fe(NH3)4]^{2+}}.}\vspace*{5pt}
  \label{tab:FeN4-geometry}
  \setlength{\tabcolsep}{2pt}
  \begin{tabular*}{0.9\textwidth}{@{\extracolsep{\fill}} c S[table-format=-1.10] S[table-format=-1.10] S[table-format=-1.10] @{\hspace{1.5em}} c S[table-format=-1.10] S[table-format=-1.10] S[table-format=-1.10]}
    \toprule
    {Atom} & {$x$} & {$y$} & {$z$} & {Atom} & {$x$} & {$y$} & {$z$} \\
    \midrule
\ce{Fe} &  0.0000 &  0.0000 &  0.0000 & \ce{N}  &  0.0000 &  2.0100 &  0.0000 \\
\ce{N}  &  2.0100 &  0.0000 &  0.0000 & \ce{H}  &  0.0000 &  2.6100 &  0.8000 \\
\ce{H}  &  2.6100 &  0.0000 &  0.8000 & \ce{H}  &  0.0000 &  2.6100 & -0.8000 \\
\ce{H}  &  2.6100 &  0.0000 & -0.8000 & \ce{H}  & -0.8000 &  2.6100 &  0.0000 \\
\ce{H}  &  2.6100 &  0.8000 &  0.0000 & \ce{N}  &  0.0000 & -2.0100 &  0.0000 \\
\ce{N}  & -2.0100 &  0.0000 &  0.0000 & \ce{H}  &  0.0000 & -2.6100 &  0.8000 \\
\ce{H}  & -2.6100 &  0.0000 &  0.8000 & \ce{H}  &  0.0000 & -2.6100 & -0.8000 \\
\ce{H}  & -2.6100 &  0.0000 & -0.8000 & \ce{H}  &  0.8000 & -2.6100 &  0.0000 \\
\ce{H}  & -2.6100 & -0.8000 &  0.0000 &         &         &         &         \\
    \bottomrule
  \end{tabular*}
\end{table}
\begin{table}
  \centering
  \caption{Optimized Cartesian coordinates (\si{\angstrom}) of \oxoMn{}.}\vspace*{5pt}
  \label{tab:oxoMn-geometry}
  \setlength{\tabcolsep}{2pt}
  \begin{tabular*}{0.9\textwidth}{@{\extracolsep{\fill}} c S[table-format=-1.10] S[table-format=-1.10] S[table-format=-1.10] @{\hspace{1.5em}} c S[table-format=-1.10] S[table-format=-1.10] S[table-format=-1.10]}
    \toprule
    {Atom} & {$x$} & {$y$} & {$z$} & {Atom} & {$x$} & {$y$} & {$z$} \\
    \midrule
    \ce{Mn} & -0.0128264866 & -0.0002779912 & -0.1091894544 & \ce{H}  &  0.8908035853 &  3.6241374839 &  1.0039698993 \\
    \ce{O}  & -1.1037647823 &  1.0246639195 & -1.2821459116 & \ce{H}  &  2.5024254412 & -2.4854165868 &  1.5399007635 \\
    \ce{O}  & -0.4319309790 & -1.5910632792 & -1.0423314985 & \ce{H}  &  2.7858100041 &  1.4098930461 &  0.8661311665 \\
    \ce{N}  &  0.7243918073 &  1.6205280457 &  0.7709992473 & \ce{H}  &  1.1490336627 & -0.0652673733 &  2.9502674019 \\
    \ce{N}  &  1.3008767173 & -0.9287768925 &  1.0635839133 & \ce{H}  &  2.0638511115 &  2.0769335299 &  2.3353274192 \\
    \ce{C}  & -1.2886186814 &  2.2834208702 & -1.2181338272 & \ce{H}  &  2.7993768711 & -0.4085469833 &  2.4534782102 \\
    \ce{C}  & -0.7050865946 &  3.1821062059 & -0.3604271375 & \ce{H}  & -1.9892347032 &  2.6686747336 & -1.9454961675 \\
    \ce{C}  &  0.3534920334 &  2.8195308082 &  0.5207059889 & \ce{H}  & -0.0986239324 & -3.4441353556 & -1.6764363233 \\
    \ce{C}  &  0.2177419816 & -2.6945497951 & -0.9662726987 & \ce{Cl} &  1.8821942342 &  0.3098852687 & -1.6599082362 \\
    \ce{C}  &  1.2181378357 & -3.0250490115 & -0.0961995911 & \ce{O}  & -1.1395010178 & -0.1703680057 &  0.9263551082 \\
    \ce{C}  &  1.7180430400 & -2.1229929449 &  0.8899899444 & \ce{H}  & -0.9312248148 &  4.2251864392 & -0.4696082573 \\
    \ce{C}  &  1.9393113446 &  1.3525171542 &  1.5367699106 & \ce{H}  &  1.6695378989 & -3.9953914470 & -0.1633815245 \\
    \ce{C}  &  1.8333722479 & -0.0604962603 &  2.1055467704 &         &               &               &               \\
    \bottomrule
  \end{tabular*}
\end{table}
After collecting the hardware samples, we applied both Cartesian and spin-adapted post-processing, including configuration recovery, to the same dataset. For the Cartesian subspace, we generated three batches of $6\times 10^3$ samples each by sub-sampling the original dataset and performed five rounds of configuration recovery. Since $N_\alpha = N_\beta$ holds only for the singlet state, we did not construct the symmetrized Cartesian subspace, thereby ensuring consistency across spin states.

For the spin-completed post-processing, we generated three sub-sampled batches of $2 \times10^4$ samples each and performed $50$ rounds of configuration recovery. The difference in batch sizes reflects the substantially smaller subspaces generated by the spin-adapted approach for a given number of samples. This allowed us to increase the batch size beyond $6\times 10^3$ samples without exceeding memory constraints, whereas larger Cartesian batches were computationally infeasible due to memory limitations. 
The number of configuration recovery rounds also differed between the two approaches: Cartesian post-processing typically converged within a few iterations, whereas the spin-completed energies continued to improve over several tens of iterations. 
Despite the larger number of iterations, the smaller subspaces kept the total computational time lower than or comparable to that of the Cartesian workflow.
Selected subspace diagonalization is performed through the SBD diagonalizer~\cite{shirakawa2025}.
In the case of the spin-completed subspaces we applied the spin projection of \cref{app:subspacediag}.

\end{document}